%% file: main.tex
\documentclass[electronic,preprint]{vgtc}   

\usepackage{times}                     
\usepackage{mathptmx}                  

\usepackage{graphicx}
\usepackage{booktabs}
\usepackage{amsmath}
\usepackage{amssymb}
\usepackage{microtype}
\usepackage{url}
\usepackage{cite}
\usepackage{balance}
\usepackage{xspace}
\usepackage{xcolor}
\usepackage{subcaption}
\usepackage{pifont}

\newcommand{\iDotE}{i.\,e.,\xspace}
\newcommand{\eDotG}{e.\,g.,\xspace}

\newcommand{\tv}{\textsc{Tribe~v2}\xspace}

\definecolor{cmBlue}{HTML}{009988}
\newcommand{\cm}{\textbf{\color{cmBlue}\ding{51}}\xspace}
\definecolor{xRed}{HTML}{cc3311}
\newcommand{\nox}{\textbf{\color{xRed}\ding{55}}\xspace}

\newcommand{\surface}{\texttt{Surface}~\includegraphics[height=1.5ex]{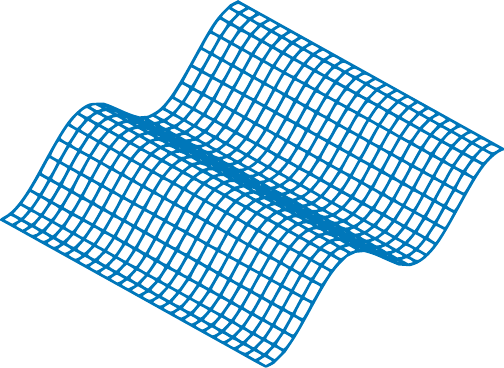}\xspace}
\newcommand{\bubble}{\texttt{Bubble}~\includegraphics[height=1.5ex]{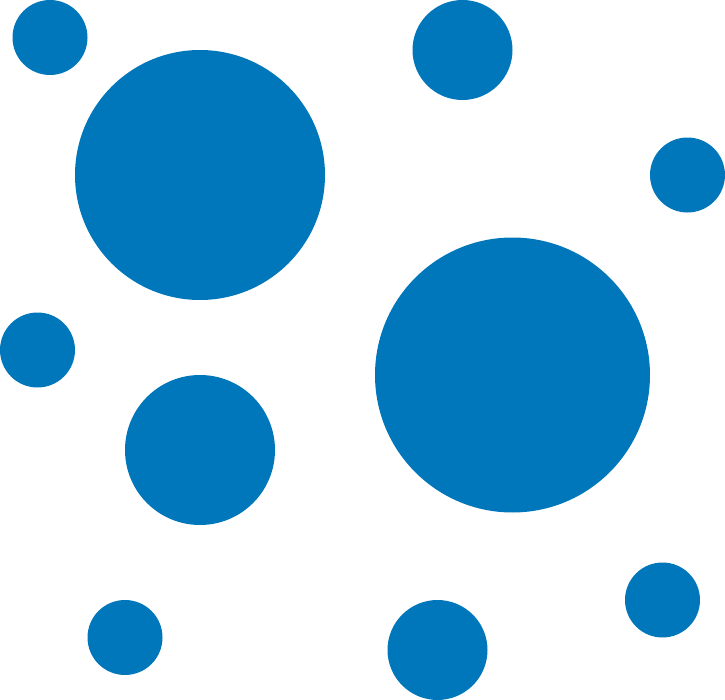}\xspace}

\onlineid{7342}
\vgtccategory{Replication}
\vgtcinsertpkg
\preprinttext{To appear in the VISxVISION Workshop at IEEE VIS 2026.}

\title{Can a Neural Encoding Model Replicate an fMRI Visualization Study?}

\author{Erfan Nasirzadeh Orang\thanks{e-mail: enorang@student.ysu.edu}
\and Zack While\thanks{e-mail: zwhile@ysu.edu}}
\affiliation{\scriptsize Department of Computer Science and Information Systems \\ Youngstown State University}

\abstract{
    \input{0-abstract.tex}
}


\begin{document}

\firstsection{Introduction}
\maketitle

\input{1-introduction}
\input{2-related-work}
\input{3-methodology}
\input{4-data-analysis-and-results}
\input{5-discussion}
\input{6-future-work-conclusion}

\balance
\bibliographystyle{abbrv-doi}
\bibliography{main}

\acknowledgments{
    This work used \textit{high performance computing} (HPC) resources provided by the Ohio Supercomputer Center~\cite{osc1987ohio}. 
    Additionally, the authors acknowledge that Claude's \textit{Opus~4.8} and \textit{Opus~5} large language models~(LLMs) were used during the process of adjusting the \tv model's provided code examples to run on our specific dataset and analyze the results for our two experiments.  We then carefully verified that the code worked as intended, and we take full responsibility for any results that are derived from such code.
}

\end{document}

%% file: 0-abstract.tex
Most knowledge of graphical perception comes from behavioral studies. Understanding from a neural perspective is much more limited due in part to neuroimaging studies' expensiveness and difficulty to conduct. In this paper, we evaluate whether Meta's \tv neural encoding model can recover neural contrasts from a visualization fMRI study.  Specifically, we evaluate \tv through a conceptual replication of the visualization-viewing component of a prior comparison of Bubble charts and three-dimensional Surface charts in color and grayscale.  We generate TRIBE-predicted cortical responses for the original stimuli and compare the resulting contrasts with those reported in the human study. The model reproduced the direction of 11 of 14 reported cortical effects, with agreement concentrated in visual-processing regions. This agreement characterizes the model's alignment with the prior human-generated fMRI results rather than independently confirming them.  We discuss the limitations encountered when working with this model for in-silico replication and hope to encourage future work exploring this new avenue for neuroimaging studies in visualization.  Supplemental materials are available at \url{https://osf.io/8a96x/}.


%% file: 1-introduction.tex
Data encoding can noticeably affect how people interpret visualizations.   Graphical perception studies have shown that different visual encodings vary in how accurately and efficiently viewers extract information from data, motivating decades of behavioral evaluation in visualization research~\cite{cleveland1984,hu2025motion}. Building on that body of work, some visualization research has sought to explain such differences by grounding their work in the perceptual and cognitive processes underlying visual analysis~\cite{ware2013information,card1999readings}.  Nevertheless, empirical visualization research has primarily examined design through the lens of user performance; while such studies can indicate which encodings are effective, they provide only indirect insight into the underlying neural processes.  This additional perspective would provide a more comprehensive view of how design choices impact interpretation and insight generation that cannot be achieved solely by metrics such as accuracy and response time~\cite{anderson2012}.  Studying those processes directly requires neuroimaging techniques such as fMRI, which depend on expensive facilities and controlled laboratory experiments~\cite{dimoka2012conduct}. 

Walden et al.~\cite{walden2018} demonstrated the value of direct neural measurement by comparing responses to two different visual encodings of the same multidimensional dataset. \bubble charts can encode values through horizontal and vertical position, mark size, and color, representing each observation as a discrete mark that must be read individually. Three-dimensional \surface charts can encode the same values as a continuous shaded terrain in space, presenting the data as a single object that may be processed more holistically. They posited that the distinction could theoretically matter since terrain-like structure may invoke perceptual machinery that discrete symbolic marks do not, which the authors framed through evolutionary theory in their study. They found differences in both behavioral performance and fMRI activations between the two, with stronger activation for \surface charts in ventral regions of the brain; 
this provided neural evidence consistent with the authors' evolution-based hypothesis beyond observed behavioral differences.

Recent advances in neural encoding models provide a potential alternative to arranging and funding traditional fMRI studies.  Meta's \tv neural encoding model was trained on more than 1,000 hours of paired fMRI and multimodal stimuli (video, text, and audio) collected across 720 subjects.  It predicts whole-brain surface activation without requiring subject-specific fMRI scans at inference time.  However, it remains unclear whether the model generalizes to static stimuli and highly artificial images such as computer-generated data visualizations.  Evaluating an alternative to direct neural measurement requires previously reported human fMRI results for comparison.  Thus, we evaluated \tv through a conceptual replication of the visualization-viewing component of Walden et al.'s study~\cite{walden2018}, investigating whether the model's predicted activations recover the neural contrasts reported for \bubble and three-dimensional \surface charts.  This aligns with work promoting replication studies~\cite{wilson2014replichi,kosara2018skipping} and critiquing the validity of the field's empirical underpinnings due to a lack of replication~\cite{kosara2016empire}.

We used \tv to simulate responses to the stimuli from both of their experiments, conducting the same subtractive analysis and comparing those results to the originals, grouped by brain region.  We observed general directional agreement with the prior study, reproducing the direction of most reported cortical effects despite methodological compromises imposed by the model's limitations.  The primary contributions of this work are: (1) the first known evaluation of a predictive neural encoding model through a conceptual replication of an fMRI-based visualization study, (2) an exploratory analysis of the results and their implications for \tv's current utility in visualization research, and (3) a discussion of methodological considerations and practical guidance for visualization researchers interested in using \tv.

%% file: 2-related-work.tex
\section{Related Work}

\noindent\textbf{Behavioral graphical perception.} 
Cleveland and McGill~\cite{cleveland1984} provided one of the earliest studies ranking perceptual tasks primarily by reading accuracy while also considering response time; Heer and Bostock~\cite{heer2010} later reproduced that work at scale using crowdsourcing. Szafir expanded on the concept of evaluating participants by considering both user tasks and chart types~\cite{szafir2018}.  A growing body of empirical work has diversified the ways that the field evaluates user performance and effective design, including memorability~\cite{borkin2016beyond}, interactivity~\cite{wall2019markov}, and trustworthiness~\cite{elhamdadi2024vistrust}.\\

\noindent\textbf{Grounding visualization in vision science.} 
Others argue for greater consideration of how the visual system works instead of solely focusing on behavioral measures~\cite{rensink2014,schonborn2024}. Scaife and Rogers~\cite{scaife1996} proposed the term \textit{external cognition}, emphasizing the interaction between external graphical representations (\eDotG visualizations) and internal cognitive representations (\eDotG mental models), and Anderson~\cite{anderson2012} argued for evaluating visualizations with cognitive and physiological measures rather than performance alone. Neural responses provide a direct means of studying the brain processes engaged during visualization, yet they are uncommon in the literature, since collecting such information requires access to expensive imaging facilities and controlled laboratory studies~\cite{dimoka2012conduct}.\\

\noindent\textbf{Neural measurement of visualization.}
Walden et al.~\cite{walden2018} and the earlier Safi et al.~\cite{safi2015} placed participants in an fMRI scanner while they answered multiple-choice questions about \bubble charts and three-dimensional \surface charts, using the resulting fMRI responses to characterize how the two types of visualizations are neurally processed.  Li et al.~\cite{livental2014} also utilized an fMRI scanner to compare responses to bar charts and line charts, reporting greater involvement of ventral-stream regions associated with recognition and interpretation than dorsal-stream regions associated with spatial and action-oriented processing.  Meanwhile, both Anderson~\cite{anderson2012} and Idesis et al.~\cite{idesis2026} recently reviewed methods for monitoring neural signals, with the latter considering their future use in generating data visualizations that adapt to a user's needs in real time.  Unlike these studies, which directly measured participants with fMRI, we evaluate whether a predictive neural encoding model can approximate the neural responses observed in such experiments.\\

\noindent\textbf{Models as proxies.}
A growing direction of research involves tasking deep learning models such as convolutional neural networks (CNNs)~\cite{haehn2018} and large language models (LLMs)~\cite{li2024,nguyen2025,poonam2026} with understanding data visualizations, often comparing their responses to human participants.  The field of human computer interaction has recently begun efforts to consider its approach toward the use of LLMs to simulate participants in human subjects studies~\cite{agnew2026workshop}; data visualization has also begun seeing studies investigating this style of synthetic participation~\cite{satkunarajan2026llms}.  
Rather than simulating subjective judgments or task performance, we use a deep learning model to predict cortical responses to specific types of visualizations.\\

\noindent\textbf{Replication in visualization and HCI.}
Extant work in visualization has provided methods of classifying replication studies based on various aspects of their design or their novelty.  Hornb{\ae}k et al.~\cite{hornbaek2014once} distinguished strict, partial, and conceptual replications according to the degree to which the original measures, manipulations, and setting were retained. Quadri and Rosen~\cite{quadri2019you} complemented this with re-evaluation, expansion, and specialization, sorting studies by the objective of their novel contribution. We use a conceptual replication as an evaluation setting for a new data-generating method.

%% file: 3-methodology.tex
\section{Methodology}
This section outlines major details regarding the study's design.  

\subsection{Conceptual Replication as Model Evaluation}\label{sec:rep-details}
Walden et al.~\cite{walden2018}, which will be referred to as \textit{the prior study}, ran two fMRI experiments comparing how people read \bubble and three-dimensional \surface charts. Participants viewed each chart alongside a multiple-choice question and answered at their own pace while neural activations were recorded.  They defined the questions as being \textit{extraction}~(answers based on visible information shown on the chart) and \textit{integration}~(answers requiring reasoning about information on the chart) tasks. The authors computed a subtractive contrast between neural responses to the two chart types and reported the resulting differences as clusters, each labeled by brain region.  Their first experiment used color charts encoding four data dimensions (\autoref{fig:colorbub} and \autoref{fig:stitch}), with each \bubble chart showing all four while the corresponding \surface chart required showing two charts with three dimensions depicted for the same data.  Following this design, we simulated responses to the same stimuli, applied the same subtractive contrast, and compared the direction of each effect with the corresponding reported cluster.

With that in mind, our objective was not to independently confirm or challenge the validity of the prior study's human fMRI findings. We instead used its published results as an empirical reference for evaluating whether \tv can reproduce known cortical contrasts elicited by data visualizations. The model, rather than the original findings, was therefore the primary object of evaluation.  Even so, our approach still reproduced several elements of the prior study's design, including its stimuli, chart comparison, subtractive contrast, and region-level results.  Using Hornb{\ae}k et al.'s classification, we therefore characterize this work as a conceptual replication, since it investigates an existing effect using a substantially different measurement system and experimental setting~\cite{hornbaek2014once}.  Under Quadri and Rosen's taxonomy, the study would be labeled as a re-evaluation: the novelty here lies in assessing whether a neural encoding model can recover results previously obtained through human fMRI~\cite{quadri2019you}. 

Our results thus do not independently confirm the prior study, as the replicated comparison instead provides a benchmark for evaluating \tv's behavior on visualization stimuli, similar to prior work that reproduced established findings to evaluate a new data-collection setting~\cite{heer2010}.  A direct replication would need to closely match the data collection conditions of the original, which is not possible due to a lack of human participants.  Additionally, the behavioral task-completion portion required responses to multiple-choice questions, which \tv cannot provide; thus, we only replicated the visualization-viewing portion of the study, striving to otherwise align with the prior study's intent within the model's limitations. 

\subsection{Stimuli}
\begin{figure*}[h]
    \centering
    \begin{subfigure}{0.19\textwidth}
        \includegraphics[width=\linewidth]{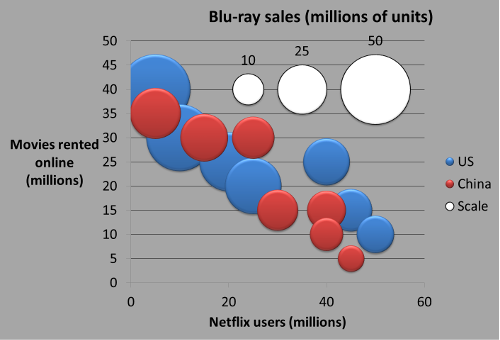}
        \caption{\bubble, Color}
        \label{fig:colorbub}
    \end{subfigure}
    \hspace{0.2em}
    \begin{subfigure}{0.36\textwidth}
        \includegraphics[width=0.95\linewidth]{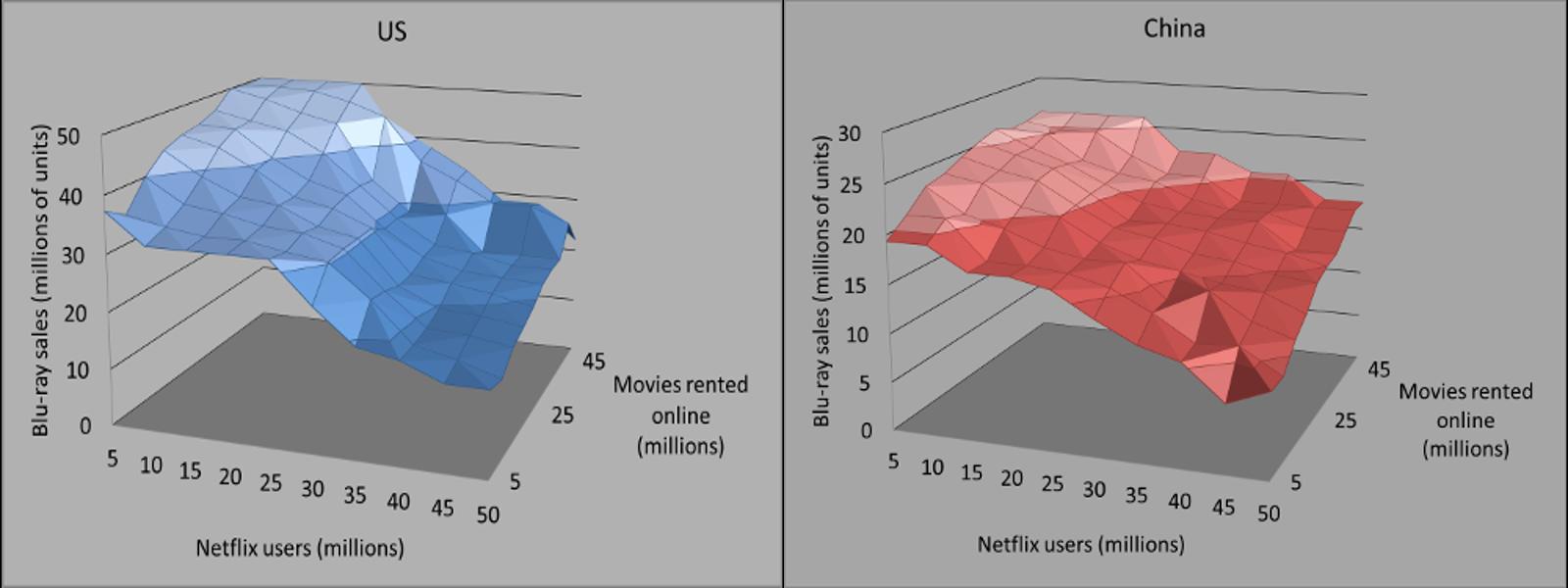}
        \caption{\surface, Color (Blue, Red), Stitched}
        \label{fig:stitch}
    \end{subfigure}
    \hfill
    \begin{subfigure}{0.178\textwidth}
        \includegraphics[width=\linewidth]{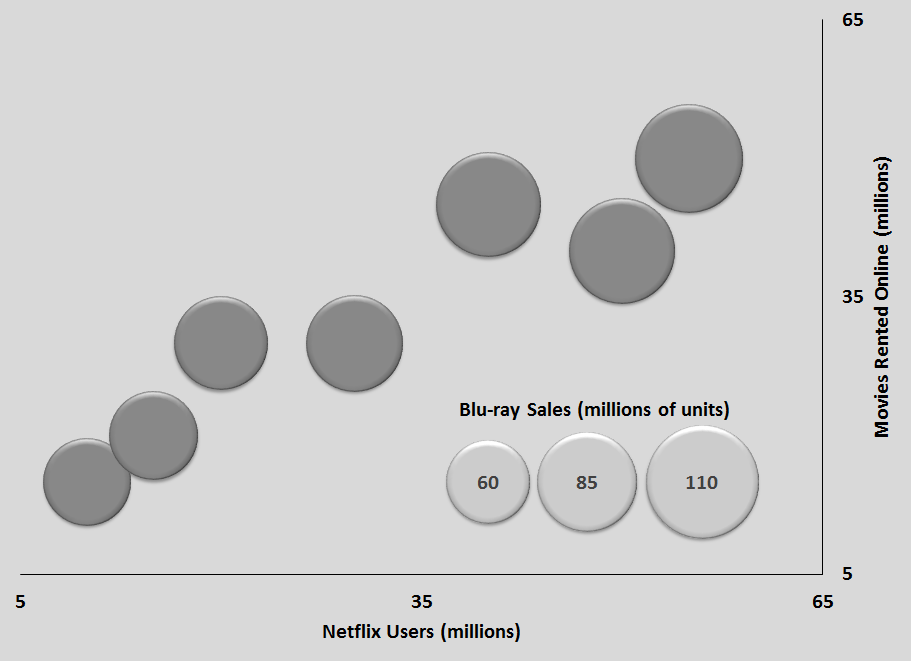}
        \caption{\bubble, Grayscale}
    \end{subfigure}
    \hspace{0.2em}
    \begin{subfigure}{0.178\textwidth}
        \includegraphics[width=\linewidth]{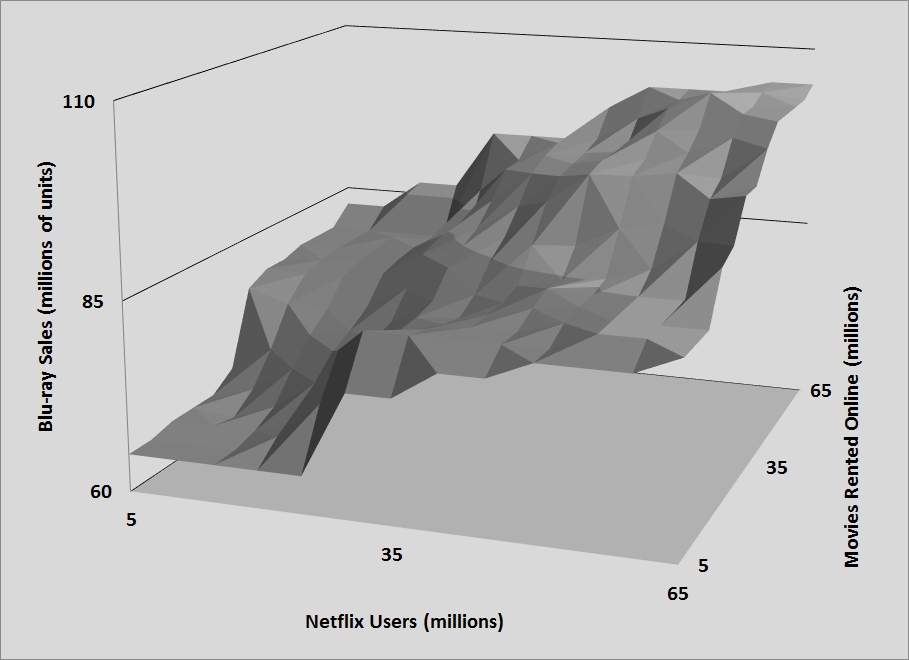}
        \caption{\surface, Grayscale}
    \end{subfigure}
    \caption{Example stimuli from Experiment 1 (a-b) and Experiment 2 (c-d), which were provided by the authors of the prior study~\cite{walden2018}.}
    \label{fig:stimuli}
\end{figure*}

We evaluated \tv using the visualization-viewing components of both experiments from the prior study after obtaining the original stimuli from the authors~\cite{walden2018}.  Experiment~1 used colored data visualizations encoding four data dimensions, consisting of three numeric features and one categorical feature. Because a \surface encodes only three dimensions, its \surface condition presents two charts side by side that each represent one category, which we stitched into one image~(\autoref{fig:stitch}). Out of concern for the impact of comparing two images instead of simply looking at one for the \surface condition, Experiment~2 used grayscale three-dimensional graphs, entirely removing the categorical feature from \surface. Each experiment consisted of 60 matched \bubble and \surface pairs (120 images total), with each pair accompanied by the same question. Following the advice of Benchetrit et al.~\cite{benchetrit2026} for working with static images, we converted each image to a 3-second silent video with the frame held constant. Images kept their native resolution (even-dimension crop only, no scaling or padding), and no audio track was added; notably, video-only input is supported by the model.  Notably, this differs from the prior study by omitting the task question, as \tv does not currently provide a straightforward mechanism for reproducing the original question-answering task.  The model thus responds only to the visualization stimulus.

We ran \tv in its default \textit{unseen-subject mode}, which predicts a single population-level response on the \texttt{fsaverage5} cortical \surface of the brain (20{,}484 vertices), generating samples at a rate of one per second (1~Hz). The model's predictions already incorporate the roughly 5-second \textit{hemodynamic delay} that requires additional exposure to get the strongest \textit{Blood-Oxygenation-Level-Dependent} (BOLD) response. Although \tv emits one prediction per second, we only report responses to the first timestep, following the single-timestep truncation described by Benchetrit et al.~\cite{benchetrit2026}.  To check for robustness, we arrived at similar results for both the color and grayscale models when truncating at time steps $t=\{0,1,2\}$, which are provided in the supplemental materials.

\subsection{Procedure}
After converting the stimuli to three-second static videos with no audio, we processed each one with the model, resulting in a set of vertices representing activations of an \textit{fsaverage5}\footnote{A standardized cortical surface format used by neuroimaging software.} cortical surface of the brain for each stimulus, stored as a NumPy array for analysis.  For each experiment's 60 matched pairs, comprising 120 images per experiment (color and grayscale), inference (prediction generation) took approximately one hour on a 32 GB NVIDIA Tesla V100 SXM2 GPU~\cite{osc1987ohio}.  The model's memory (VRAM) usage peaked at approximately 1.07 GB and remained near that level throughout.

%% file: 4-data-analysis-and-results.tex
\section{Data Analysis and Results}

Here we summarize our method of analysis as well as the results.

\subsection{Analysis Method}
For each pair, we computed a cognitive-subtraction contrast, \surface minus \bubble, at every vertex. This was the same contrast Walden et al.\ ran and the standard approach in prior fMRI graph studies~\cite{livental2014,safi2015}. We averaged across the 60 pairs and assigned each vertex to a region using the Destrieux atlas~\cite{destrieux2010}, excluding non-cortical labels.  Because the model is deterministic, the 60 pairs are 60 distinct stimuli instead of independent measurements of one effect. Averaging across them reduced variation from individual images (differences in visual density or value range) rather than noise and isolated the part of the contrast attributed to graph type.  

For each of the prior study's reported clusters, we checked whether the predicted contrast had the same direction, \iDotE whether the same chart type (\surface or \bubble) had stronger activations. Comparison was restricted to cortical clusters, since \tv's surface output does not cover the cerebellar and subcortical clusters that the prior study also reported.  Furthermore, as an additional sanity check for these results, we repeated the analysis after removing each pair's whole-brain mean, which is available in the supplemental materials.  This step removed global differences in predicted activation magnitude and tested whether the observed effects persisted as regional deviations from the brain-wide average.  Lastly, to measure uncertainty, we applied BCa bootstrapping~\cite{efron1987better} to the contrasts from the stimulus pairs, computing a 95\% confidence interval (CI) of the mean difference across the stimuli. 

\subsection{Directional Agreement as a Coarse Comparison}
We note here that direct numeric comparison between our results and the prior study is not well-defined. The prior study's $z$-scores come from variance across 20 human participants, each with measurement and physiological noise. \tv is deterministic and, in \textit{unseen-subject mode}, produces a single population-level prediction~\cite{tribev2}, so there is no measurement variance to standardize against.  We therefore report each contrast's direction (\iDotE its sign) and raw magnitude, evaluating directional agreement only within the brain regions highlighted by the prior study. Directional agreement indicates whether \tv predicts the same chart type (\surface or \bubble) to produce the stronger response within a reported region. It thus provides a coarse measure of whether the model preserves the relative ordering of the two conditions observed in the prior results.  This agreement provides evidence about \tv's correspondence with the prior results.  It does not independently confirm the existence or reliability of the original human effects, nor does it demonstrate that the model recaptured the effects' magnitude, spatial distribution, statistical reliability, or functional significance.

\subsection{Color Stimuli (Experiment 1)}
\tv matched 7 of the prior study's 9 cortical clusters~(\autoref{tab:color}).
Both fusiform clusters, left superior lateral occipital, and all four \bubble-dominant\footnote{\bubble activations higher than \surface activations}
clusters matched. On the other hand, the left inferior lateral occipital 
and left superior parietal lobule disagreed with the prior study's direction.
Some predicted differences were close to zero, including those in the left middle frontal gyrus~($-0.011$) and right anterior cingulate~($-0.017$). 
The largest absolute mean contrast was~$0.087$ in the left occipital pole.

\begin{table}[h]
\caption{Comparison of the model's predicted contrasts with the prior study's results for \textit{Experiment 1} (color). Positive values indicate higher predicted fMRI response for \surface. A~\cm symbol indicates alignment with the corresponding result of the prior study, while a~\nox symbol indicates disagreement. L stands for \textit{left} and R stands for \textit{right}. 
In the \tv repository demo, responses to a 52-second within-distribution color video with audio had a middle 95\% range of approximately $[-0.35, 0.54]$ across time steps ($\mu=0.07$).
}
\label{tab:color}
\resizebox{\columnwidth}{!}{
\begin{tabular}{llll}
\hline
\textbf{Cluster (Walden)} & \textbf{Prior Result} & \multicolumn{1}{c}{\textbf{Ours (Mean)}} & \textbf{Ours (95\% CI)} \\ \hline
L Superior Lateral Occipital & \surface & $+0.056$ \cm & $[+0.044, +0.071]$ \\
R Temporal Occipital Fusiform & \surface & $+0.041$ \cm & $[+0.033, +0.049]$ \\
L Temporal Occipital Fusiform & \surface & $+0.019$ \cm &$ [+0.011, +0.027]$ \\
L Inferior Lateral Occipital & \surface & $-0.030$ \nox & $[-0.042, -0.016]$ \\
L Superior Parietal Lobule & \bubble & $+0.038$ \nox & $[+0.031, +0.047]$ \\
L Middle Frontal Gyrus & \bubble & $-0.011$ \cm & $[-0.012, -0.010]$ \\
R Anterior Cingulate & \bubble & $-0.017$ \cm & $[-0.020, -0.015]$ \\
R Middle Precentral Gyrus & \bubble & $-0.021$ \cm & $[-0.023, -0.020]$ \\
L Occipital Pole & \bubble & $-0.087$ \cm & $[-0.099, -0.072]$ \\ \hline
\end{tabular}
}
\end{table}

\subsection{Grayscale Stimuli (Experiment 2)}
\tv matched the direction of 4 of the prior study's 5 cortical clusters (\autoref{tab:grayscale}): left superior lateral occipital ($+0.142$), right fusiform ($+0.095$), right cuneus ($-0.049$), and left middle frontal gyrus matched weakly ($+0.012$).  The right superior frontal did not match the prior study ($-0.011$). Here two clusters were close to zero (left middle frontal gyrus and right superior frontal gyrus) and the largest absolute value was the left superior lateral occipital ($0.142$).

\begin{table}[h]
\caption{Grayscale (Experiment 2). Predicted contrast at the prior study's reported cortical clusters. Positive values indicate \surface-dominance. A~\cm shows matching the results of the prior study, while~\nox shows disagreement.  L stands for \textit{left} and R stands for \textit{right}.  
In the \tv repository demo, responses to a 52-second within-distribution color video with audio had a middle 95\% range of approximately $[-0.35, 0.54]$ across time steps ($\mu=0.07$).
} 
\label{tab:grayscale}
\resizebox{\columnwidth}{!}{
\begin{tabular}{llcc}
\hline
\textbf{Cluster (Walden)} & \textbf{Prior Result} & \textbf{Ours (Mean)} & \textbf{Ours (95\% CI)} \\ \hline
L Superior Lateral Occipital & \surface & $+0.142$ \cm & $[+0.135, +0.149]$ \\
R Fusiform Cortex & \surface & $+0.095$ \cm & $[+0.089, +0.100]$ \\
L Middle Frontal Gyrus & \surface & $+0.012$ \cm & $[+0.011, +0.013]$ \\
R Superior Frontal Gyrus & \surface & $-0.011$ \nox & $[-0.013, -0.010]$ \\
R Cuneus & \bubble & $-0.049$ \cm & $[-0.052, -0.045]$ \\ \hline
\end{tabular}
}
\end{table}

\subsection{Consistency Across Experiments}
Left superior parietal was \surface-dominant in both experiments and both analyses. It is not among the prior study's Experiment~2 clusters, but the corresponding parcel in our grayscale data was also \surface-dominant ($+0.146$), matching the color result ($+0.038$) and running opposite of the prior study's \bubble direction. In total, \tv matched the direction of 11 of 14 cortical clusters.

%% file: 5-discussion.tex
\section{Discussion}
Our study evaluated \tv through a conceptual replication of selected components of Walden et al.'s experiments~\cite{walden2018}; it did not reproduce the complete human experimental procedure~(see \autoref{sec:rep-details}).  Participants in their study viewed each stimulus while answering a multiple-choice question based on an \textit{extraction} or \textit{integration} task, whereas \tv only received a visualization with no task.  Our process thus lacked the prior study's task demands, and the results must be interpreted with this distinction in mind.\\

\noindent\textbf{Representing task instructions is an open question.}
\tv is not designed to extract written questions from images and instead processes language through audio-derived text~\cite{tribev2}.  However, doing so would instead be simulating a participant \textit{hearing} a question as opposed to reading it.  The proper way to provide questions to the model is still an open question, and as of now the recommendation of \tv's authors is for audio-free videos~\cite{benchetrit2026}.  More work is needed to investigate how task instructions can be provided to neural encoding models while still aligning with human studies.\\

\noindent\textbf{Given only visual stimuli, \tv's agreement was concentrated in perceptual regions.} 
The clusters that agreed with the prior study were primarily visual and occipital regions, including middle and lateral occipital cortex, lingual gyrus, fusiform, and cuneus.  In contrast, the frontal regions examined in the prior study produced some of the smallest differences in our analysis. Because the prior study required participants to view the visualizations and answer questions, these regions' lower agreement may partially reflect task demands such as working memory, response selection, or comparison processes that were absent from our image-only stimuli.  We caution that our results should not be interpreted as demonstrating that \tv reproduces human visualization perception broadly.  The model aligned with the direction of several reported contrasts, suggesting that it may capture some perceptual distinctions between the chart types reflected in the prior study.\\

\noindent \textbf{\tv shows consistent behavior across varying stimuli.}
The perceptual match held across a change in rendering (color vs.\ grayscale), dimensionality (data with four features vs. data with three), and stimulus style (single image vs.\ stitched images). 
The agreement persisted across both experiments despite differences in color, data dimensionality, and presentation, providing some evidence that the observed agreement is robust to such changes.\\

\noindent\textbf{Not all disagreements are explained by task omission.}
One notable disagreement between our results and the prior study was the left superior parietal region, which was consistently \surface-dominant across both experiments despite showing the opposite direction in the prior study. Because parietal regions are often associated with visuospatial processing, this is not as readily explained by the absence of the question-answering task. It may, instead, reflect a limitation of the model for visualization stimuli or sensitivity to aspects of visual inputs that differs from humans.\\

\noindent\textbf{\tv predictions introduce methodological limitations.}
The model's predictions are deterministic and generate a single population response, compared to the prior study's use of $z$-scores to illustrate response magnitude. This motivated our focus on the direction of effect rather than specific numerical outcomes in the analysis.  As a result, our evaluation should instead be viewed as a qualitative comparison of agreement, as matching effect direction alone does not demonstrate that the underlying neural responses are identical to those reported in the prior study.  Mapping the prior study's voxel-based clusters onto the parcels used by \tv required name-based matching, introducing the possibility of small misalignments due to differing representations.  The stitched color \surface charts differed from \bubble stimuli in aspect ratio, which the model's fixed input resizing compresses unequally.  Static, silent images are out-of-distribution for a video-trained model, so as of now visualization stimuli may lead to less consistent results.  This may have contributed to the relatively small differences observed in some regions (\eDotG right superior frontal gyrus).  Notably, because the model only predicts neural activations on the cortical surface, it cannot capture cerebellar or subcortical responses.\\

\noindent\textbf{Visualization researchers should work with \tv's strengths.}  
Given the lack of a natural mechanism in the model for providing task instructions with the prior study's stimuli as-is, our replication focused on comparing perceptual effects and observed the strongest agreement in visual and occipital regions.  As such, we recommend using \tv as a supplementary tool for comparing perceptual impacts of visualization design choices.  Additionally, researchers should be prepared and able to convert their visualizations into static video clips.  From our study, it is unclear how well the model is suited for non-static visualization clips (\eDotG videos of interactive, animated visualizations), and this is an open question for future work to consider.  
Because the model's output values represent a population response rather than participant-level measurements, comparisons to real-world fMRI studies should generally be interpreted qualitatively. However, the outputs can still be compared numerically against other predictions generated by the model.
One must also consider \tv's computational requirements.  
Based on our observed peak usage~($\sim$1.07 GB), inference appears feasible on GPUs with substantially less memory than the 32 GB used in our experiments.  
Furthermore, since the model only predicts activations on the cortical surface, it is most appropriate for studying cortical processes. Those studying cerebellar or subcortical activity will likely require the use of more traditional methods.

%% file: 6-future-work-conclusion.tex
\section{Future Work and Conclusion}

Models like this could provide a low-cost way to screen visualization designs against a neural benchmark, although they currently appear better suited to modeling perceptual responses than task engagement.  Future work could test different methods of adding the question text to see whether the frontal clusters return.  Furthermore, researchers could replicate additional fMRI studies such as one by Li et al.~\cite{livental2014}. 

In this paper, we evaluated \tv through a conceptual replication of the visualization-viewing component of Walden et al.'s fMRI experiments~\cite{walden2018}.  The model reproduced the effect direction of 11 of 14 cortical clusters from Walden et al.'s two fMRI experiments~\cite{walden2018}. While these results do not independently confirm the original findings, they suggest that neural encoding models may be capable of recovering some perceptual distinctions observed in visualization neuroimaging studies. However, additional work is needed to fully understand their utility.  While in its infancy as a research tool, this approach shows considerable promise and warrants some consideration from the visualization field as it continues to develop.